\documentclass{aa}
\usepackage{graphicx}
 \usepackage{gensymb}
\usepackage{natbib}
\usepackage{amsmath}
\usepackage{amssymb}
\usepackage{fancyhdr}
\usepackage{latexsym}
\usepackage{color}

\begin{document} 

\title{A study of gamma ray bursts with afterglow plateau phases associated with supernovae}
\author{M.G. Dainotti\inst{1,2,3} \and S. Nagataki\inst{4} \and K. Maeda\inst{5,6} \and S. Postnikov\inst{7} \and E. Pian \inst{1,8}}
 \institute{INAF-Istituto di Astrofisica Spaziale e Fisica cosmica, c/o CNR - Area della Ricerca di Bologna, Via Gobetti 101, 40129 - Bologna, Italy \\
\and
Physics Department, Stanford University, 382 Via Pueblo Mall, Stanford, USA \email{mdainott@stanford.edu}\\
\and
Obserwatorium Astronomiczne, Uniwersytet Jagiello\'nski, ul. Orla 171, 31-501 Krak{\'o}w, Poland \email{dainotti@oa.uj.edu.pl}\\
\and
Astrophysical Big Bang Laboratory, Riken, 351-0198, Wako Saitama, Japan \email{shigehiro.nagataki@riken.jp}\\
\and
Department of Astronomy, Kyoto University, Kitashirakawa-Oiwake-cho, Sakyo-ku, Kyoto 606-8502, Japan \email{keiichi.maeda@kusastro.kyoto-u.ac.jp}\\
\and
Kavli Institute for the Physics and Mathematics of the Universe (WPI), University of Tokyo, 5-1-5 Kashiwanoha, Kashiwa, Chiba 277-8583, Japan \\
\and
Bloomington, Indiana University, USA \email{postsergey@gmail.com}\\
\and
Scuola Normale Superiore, Piazza dei Cavalieri 7, I-56126 Pisa, Italy \email{elena.pian@sns.it}\\
}

\date{Received; accepted }
\abstract
{The analysis of $176$ gamma ray burst (GRB) afterglow plateaus observed by Swift from GRBs with known redshifts revealed that the subsample of long GRBs associated with supernovae (LONG-SNe), comprising 19 events, presents a very high correlation coefficient between the luminosity at the end of the plateau phase $L_X(T_a)=L_a$ and the end time of the plateau $T^*_a$. Furthermore, these SNe Ib/c associated with GRBs also obey the peak-magnitude stretch relation, which is similar to that used to standardize the SNe Ia.}{Our aim is to investigate a category of GRBs with plateau and associated with SNe to compare the Dainotti correlation for this sample with the correlation for long GRBs for which no associated SN has been observed (hereafter LONG-NO-SNe, 128 GRBs) and to check whether there is a difference among these subsamples.}{We first adopted a nonparametric statistical method to take redshift evolution into account and to check if and how this effect may steepen the slope for the LONG-NO-SNe sample. This procedure is necessary because this sample is observed at much higher redshift than the GRB-SNe sample. Therefore, removing selection bias is the first step before any comparison among samples observed at different redshifts could be properly performed. Then, we applied the T-student test to evaluate a statistical difference between the slopes of the two samples.}{We demonstrate that there is no evolution for the slope of the LONG-NO-SNe sample and no evolution is expected for GRBs observed at small redshifts such as those present in the LONG-SNe sample. The difference between the slope of the LONG-NO-SNe and the slope of LONG-SNe, i.e., those with firm spectral detection of SN components, is statistically significant ($P=0.005$).}{This possibly suggests that, unlike LONG-NO-SNe, LONG-SNe with firm spectroscopic features of the associated SNe might not require a standard energy reservoir in the plateau phase. Therefore, this analysis may open new perspectives in future theoretical investigations of the GRBs with plateau emission and that are associated with SNe.}

\maketitle 
\section{Introduction}
Gamma ray bursts (GRBs), which are the farthest sources seen up to redshift $z=9.46$ \cite{Cucchiara2011}, might be powerful cosmological tools. Gamma ray bursts have been traditionally classified as short ($T_{90}<2 s$, where $T_{90}$ is the time when the burst emits between $5\%$ and $95\%$ of its isotropic emission) and long ($T_{90}>2s$), although further studies revealed the existence of short GRBs with extended emission, hereafter GRB-SE \cite{nb06,nb2010}. Moreover, long GRBs have been divided into normal and low luminosity (llGRBs) classes \cite{Nakar2015}. Here, we investigate both regular luminosity long GRBs and GRBs found to be associated with SNe (GRB-SNe) to determine whether there is a subset showing standard candle properties.

Notwithstanding the variety of GRB peculiarities, some common features may be identified by looking at their light curves. Swift satellite observations of GRBs have made a crucial breakthrough in this field; these observations have revealed a more complex behavior of the afterglow emission \cite{OB06,Nousek2006,Zhang2006,Sak07} than in the pre-Swift era. Afterglow light curves can be divided into two, three, and even perhaps more segments. The second segment, when it is flat, is called plateau emission.
In addition, Willingale at al. 2007, hereafter W07, showed that Swift prompt and afterglow light curves may be fitted by the same analytical expression. This provides the opportunity to search for similar properties that would help us to understand if GRBs are standard candles. 

Within this context, Dainotti et al. (2008, 2011a, 2015a) proposed to standardize GRBs using afterglow features, such as the anti-correlation between $L_a$, the isotropic X-ray luminosity at the time $T_a$, and the rest-frame time at the end of the plateau phase, $T^*_a$ (where $*$ denotes the rest-frame quantities), hereafter called Dainotti relation. 
Later, Dainotti et al. (2013a) demonstrated that the Dainotti relation has an intrinsic slope $b=-1.07_{-0.14}^{+0.09}$ and this finding has an important implication on its possible physical explanation which may imply that a fixed energy reservoir powers the plateau. This possibility has been explored in the context of the fallback mass surrounding the black hole, according to the scenario proposed by Cannizzo \& Geherls (2009) and Cannizzo et al. (2011). Also additional theoretical interpretations have been explored to explain the Dainotti correlation, for example the slope $\approx -1$ can be derived from the spinning of a newly born magnetar \cite{Dall'Osso,Rowlinson2010,Obrien2012,Nemmen2012,Rowlinson2013,Rowlinson2014,Rea2015}. However, there are several models, such as the photospheric emission model \cite{Ito2014}, for which the Dainotti relation has not yet been tested. 

It is important to discriminate among subsamples, which are observationally different, to validate theoretical models because this difference may be due to diverse emission mechanisms \cite{Dainotti2010} and therefore can influence the use of the Dainotti relation as a cosmological tool \cite{Cardone09,Cardone2010,Postnikov2014,Dainotti2013b}. The problem of selecting homogeneous samples in terms of similar observational properties usually helps to reduce the scatter of correlations \cite{Dainotti2010,Dainotti2011a,Delvecchio2016} and it is a general problem that can be equally applied to prompt correlations \cite{Yonekotu2004,amati09,Dainotti2016b} and prompt-afterglow correlations \cite{Dainotti2011b,Dainotti2015b}. Very recently, Dainotti et al. (2016a) have combined the two intrinsic correlations: the $L_a-T^{*}_a$ correlation and the prompt-afterglow correlation, $L_{peak}-L_a$, where $L_{peak}$ (erg $s^{-1}$) is the peak prompt emission luminosity. We obtained a 3D correlation that is much tighter for long GRBs, for which SNe have not been seen. We excluded X-ray flashes (XRFs) from this sample; X-ray flashes are bursts that usually have a ratio between prompt X-ray fluence and gamma ray fluence $\ge 1$.

With this issue in mind, we focused on the updated sample of 176 GRBs with known redshift and observed plateau emission and we looked for a subset with high degree of correlation in the luminosity-time space and associated with SNe. It is the first time that such an investigation of the Dainotti relation has been performed for GRBs associated with SNe. 

In the present analysis, we use the nomenclature of LONG-SNe and LONG-NO-SNe just for simplicity, since there could be SNe associated with most GRBs, which we did not observe because no sensitive search was possible at that time or no useful upper limits to the presence of a possible SN in the optical afterglow light curve had been derived \cite{Melandri2014}.

However, we have counter examples already (although one example could indeed be a short GRB) in which we are able to put very stringent limit on the lack of a supernova emission. For example in the case of GRB LONG-NO-SNe, 060505 (Ofek et al. 2006), using the Hubble Space Telescope, placed an approximate upper limit on the mass of the radioactive nickel $56$ produced in the explosion, $M_{56Ni}$ $\leq 2 \times 10^{-4}$ M, assuming no extinction. It was noted that the faintest core collapse supernovae known to date ejected about $(2-8) \times 10^{-3}$ $M_{56Ni}$ \cite{Pastorello2004}. 

Fynbo et al. (2006b) suggested that this GRB may belong to a new emerging group of long duration GRBs without supernovae. The existence of such a class of GRBs was already mentioned in the past \cite{Mukherjee1998,Horvath2002} based on the analysis of the GRB duration distribution.
A third group was also discussed by Gal-yam et al. (2006) and Della Valle et al. (2006).
From these observations, it seems that the scenario in which long-duration, soft-spectrum GRBs (LONG-SNe) are accompanied by massive stellar explosions \cite{WB2006} requires additional explanation for the above events. In summary, this observational panorama is suggestive of the fact that there may be two types of LONG-GRBs with and without SNe. 

Therefore, it is worthwhile to investigate a reasonable distinction in these categories to better clarify such a debated issue. Since the Dainotti relationship is connected to the physics of the GRB and has been used as a model discriminator, it is worth asking whether there is evidence that this relation is significantly different for GRBs with and without SNe.

In \S \ref{LTupdates} we describe the data analysis, and in \S \ref{standard set} we divide the total sample into categories, such as LONG-SNe, LONG-NO-SNe, SE, and XRFs. In \S \ref{EP method} we discuss the motivation of applying the Efron and Petrosian (1992) method, hereafter EP. In \S \ref{standardizable candle} we discuss properties of our sample SNe associated with GRBs and their link to the peak magnitude-stretch relation of the SNe. In \S \ref{Conclusions} we present summary and main conclusions. In Appendix A we show how we remove the biases affecting the total LONG-NO-SNe sample with the EP method, thus obtaining a sample that is independent of the redshift evolution and selection effects related to GRB instrumental threshold. In Appendix B we compare the properties of the GRBs associated with SNe, for which strong spectroscopical evidence of SNe is present (seven GRBs), with the properties of seven GRBs observed at small redshift that belong to LONG-NO-SNe category in order to pinpoint whether there are features of prompt emission of the GRB-SNe that are distinctive of this class.

\section{Data analysis}\label{LTupdates}
We analyzed the sample of all (176) GRB X-ray plateau afterglows, detected by {\it Swift} from January 2005 up to July 2014 with known redshifts, both spectroscopic and photometric, available in Xiao \& Schaefer (2009), on the Greiner web page \footnote{http://www.mpe.mpg.de/~jcg/grbgen.html}, and in the Circulars Notice arxive (GCN). We exclude redshifts for which there is only a lower or an upper limit in their determination. The redshift range of our sample is $(0.033, 9.4)$. We include all the X-ray plateaus for which the afterglow light curves can be fitted by the Willingale et al. (2007), hereafter W07, phenomenological model. 
The W07 model proposed the following functional form:
\begin{equation}
f(t) = \left \{
\begin{array}{ll}
\displaystyle{F_i \exp{\left ( \alpha_i \left( 1 - \frac{t}{T_i} \right) \right )} \exp{\left (
- \frac{\tau_i}{t} \right )}} & {\rm for} \ \ t < T_i \\
~ & ~ \\
\displaystyle{F_i \left ( \frac{t}{T_i} \right )^{-\alpha_i}
\exp{\left ( - \frac{\tau_i}{t} \right )}} & {\rm for} \ \ t \ge T_i \\
\end{array}
\right .
\label{eq: fc}
\end{equation}
for both the prompt (the index `i=\textit{p}') $\gamma$\,-\,ray and initial X -ray decay and for the afterglow (``i=\textit{a}") modeled so that the complete light curve $f_{tot}(t) = f_p(t) + f_a(t)$ contains two sets of four parameters $(T_{i},F_{i},\alpha_i,t_i)$. The transition from the exponential to the power law occurs at the point $(T_{i},F_{i}e^{-\tau_i/T_i})$ at which the two functional sections have the same value and gradient. The parameter $\alpha_{i}$ is the temporal power law decay index and the time $\tau_{i}$ is the initial rise timescale.
In previous papers, such as W07 and Dainotti et al. (2008,2010), the {\it Swift} Burst Alert Telescope (BAT)+ X-Ray Telescope (XRT) light curves of GRBs were fitted to Eq. (\ref{eq: fc}) assuming that the rise time of the afterglow, $\tau_a$, started at the time of the beginning of the decay phase of the prompt emission, $T_p$, namely $\tau_a=T_p$. Here we leave $\tau_a$ free to vary. We exclude the cases that are not fit well by the W07 model, namely when the fitting procedure fails or when the determination of confidence interval in 1$\sigma$ does not fulfill the Avni 1976 prescriptions; for more details see the xspec manual \footnote{http://heasarc.nasa.gov/xanadu/xspec/manual/\newline XspecSpectralFitting.html)}. For a proper evaluation of the error bars, the Avni prescriptions require the computation in the 1$\sigma$ confidence interval for every parameter varying the parameter value until the $\chi^2$ increases by a certain value above the minimum (or the best-fit) value. These rules define the amount that the $\chi^2$ is allowed to increase, which depends on the required confidence level and on the number of parameters whose confidence space is being calculated.
 
This sample is an update of that presented in Dainotti et al. (2015a) and, for the first time, it takes into account a detailed analysis of the Dainotti relation for the GRBs-SNe compared to the LONG-NO-SNe, SE, and XRFs.  \newline
We compute the luminosity $L_a$ in the {\it Swift} XRT band pass, $(E_{min}, E_{max})=(0.3,10)$ keV as follows:
\begin{equation}
L_a= 4 \pi D_L^2(z) \, F_X (E_{min},E_{max},T_a) \cdot \textit{K},
\label{eq: lx}
\end{equation}
where $D_L(z)$ is the GRB luminosity distance for the redshift $z$, computed assuming a flat $\Lambda$CDM cosmological model with $\Omega_M = 0.291$ and $h = 0.697$, $F_X$ is the measured X-ray energy flux in (${\rm erg/cm^2/s}$), and  \textit{K} is the \textit{K} correction for cosmic expansion \cite{B01}.
The light curves are taken from the Swift web page repository, $http://www.swift.ac.uk/burst_analyser/$ and we followed the Evans et al. (2009) approach for the evaluation of the spectral parameters.
We derived the normalization, $a$, and slope, $b$, of the Dainotti relation for the distributions of the all analyzed subsamples (see \S \ref{standard set}) using the D'Agostini (2005) method fitting procedure.

\begin{figure}[!tbh]
\includegraphics[width=9cm]{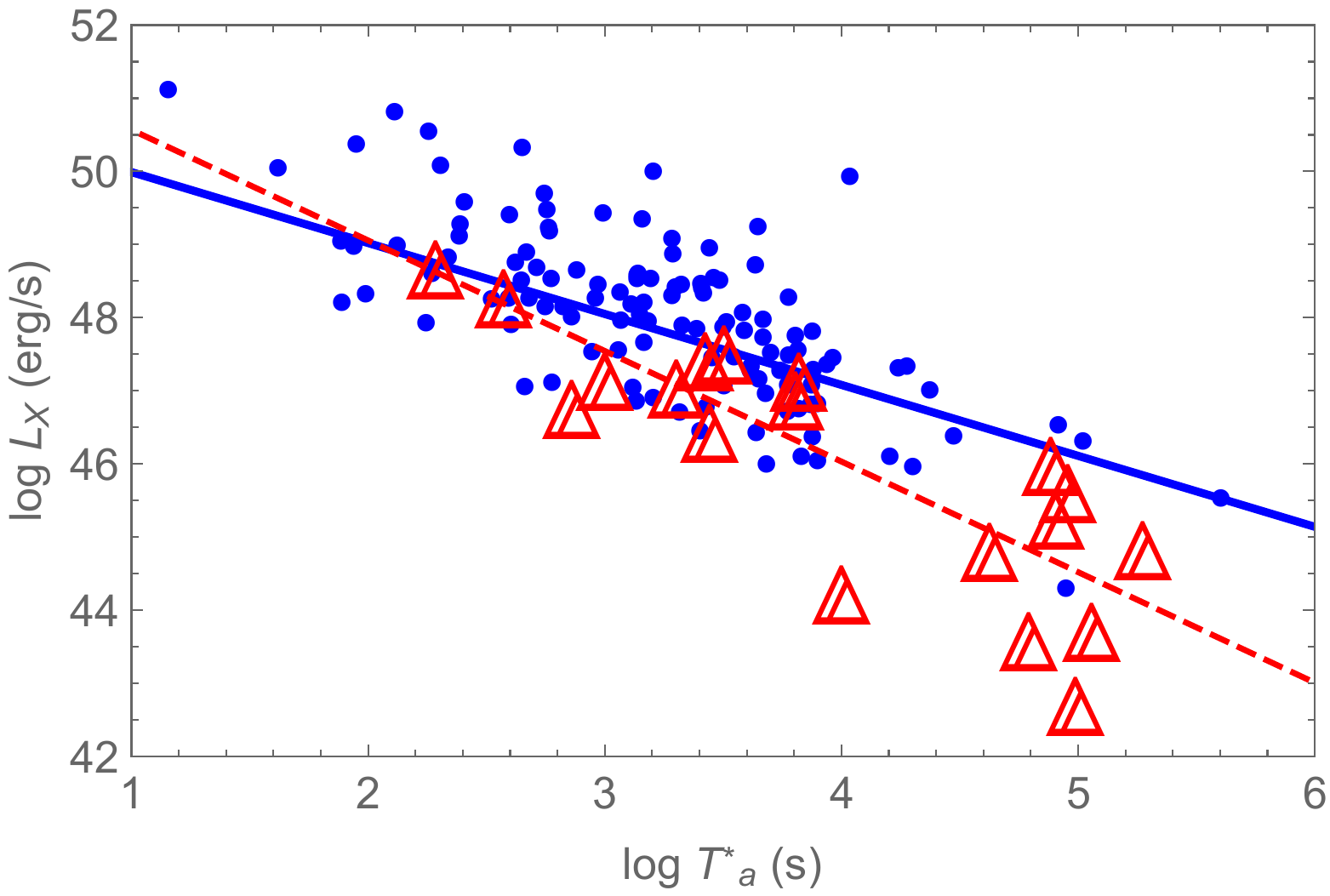}
\includegraphics[width=9cm]{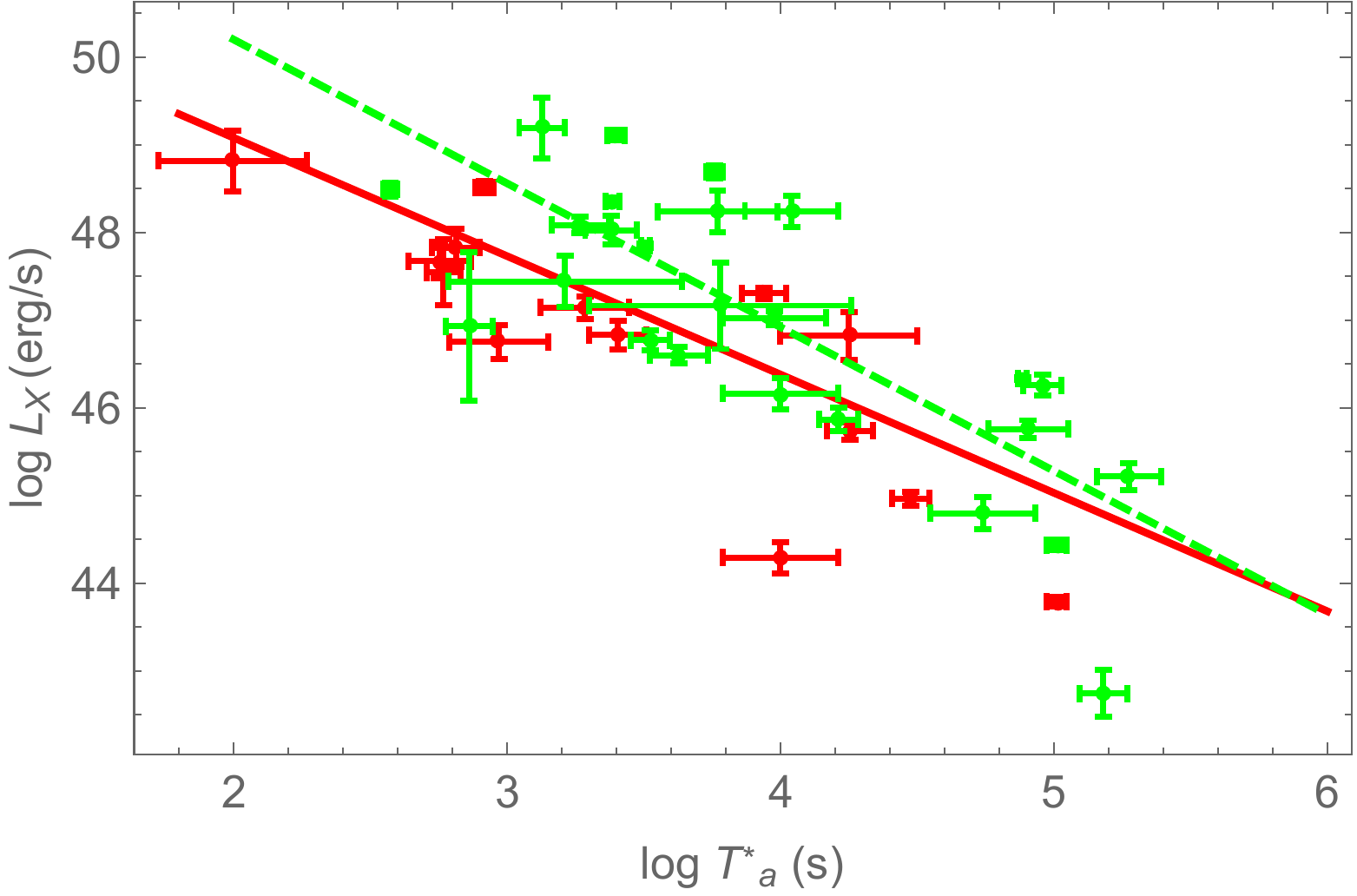}
\caption{Analyzed $\log L_X$ vs. $\log T_a^{*}$ distributions. Upper panel: LONG-NO-SNe 128 GRBs (blue points fitted with a solid blue line) and the 19 events from LONG-SNe (red empty triangles) fitted with a red dashed line. Lower panel: XRFs (25) GRBs (green points fitted with a dashed green line), and SE (16) (red points) fitted with a red solid line}
\label{fig1}
\end{figure}

\begin{figure}
\includegraphics[width=9.cm]{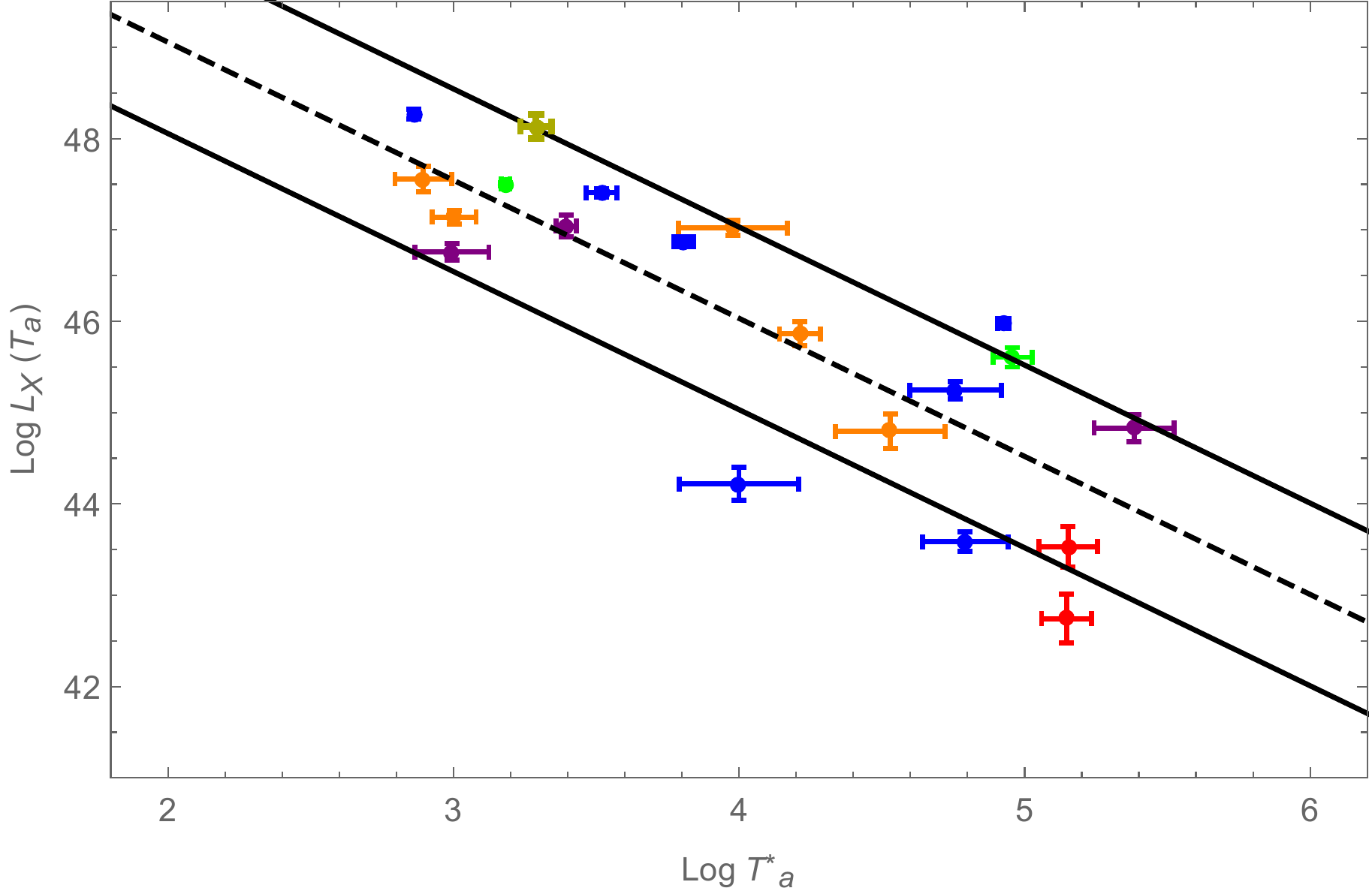}
\caption{Analyzed $\log L_X$ vs. $\log T_a^*$ distributions for LONG-SNe divided into colors depending on the category described in Table 2. Category A: red points; B: orange; C: green; D: purple; and E: blue. The dark yellow point represents a low-redshift GRB, GRB 060505, for which the SN was not seen associated with GRB; see discussion in the text. The two solid lines represent the 1$\sigma$ intrinsic scatter of the Dainotti relation, while the dashed line is the best-fit line computed with the D'Agostini method. 
\label{fig2}}
\end{figure}

\section{The Dainotti relation in GRB subsamples}\label{standard set}

Our aim is to find common trends among the selected $176$ GRB light curves ($160$ LONG+$16$ Short with extended emission), all observed by Swift. Within the LONG sample (160 GRBs) we analyze the subsamples of $128$ LONG-NO-SNe and $19$ LONG-SNe presented as blue points and red empty triangles, respectively, in the upper panel of Fig. \ref{fig1}. Within the LONG sample we also analyze $25$ XRFs (see green points in the lower panel of Fig. \ref{fig1}); 12 GRBs are common in the LONG-SNe and XRFs samples; 16 short GRBs with extended emissions (SE) are presented in the lower panel of Fig. \ref{fig1}. 

\subsection{GRBs associated with SNe}
Since the category of associated GRB-SNes and obeying the Dainotti relation has not yet been discussed so far, we address this investigation here.
We present the fitted correlation slope, $b,$ and its error, $\delta_b$, the normalization, $a$, and its error, $\delta_a$, the $\rho_{LT}$, and the probability $P$ for the entire analyzed subsample; see Table \ref{Table1} and Figs. \ref{fig1} and \ref{fig2}. 
For all the subsamples, the Spearman correlation coefficient, $\rho_{LT}$, is always greater than $0.7$ and the probability that the fitted correlation occurred randomly in an uncorrelated data set is $\emph{P} \le 10^{-3}$ \citep{Bevington}, thus confirming the existence of these correlations. The difference between the slopes of LONG-NO-SNe and SE is not statistically significant ($P=0.20$) possibly because
of the paucity of the sample of SE, while the difference between the slopes of the LONG-NO-SNe and XRFs is significant only at the $6\%$ level because the probability is $P=0.06$. This difference is in part also due to the presence of GRB-SNe in this sample. Also the difference between the LONG-NO-SNe and the total sample of SNe is only significant at the $11\%$ level. However, the main focus of this investigation is to address the difference between the LONG-NO-SNe and LONG-SNe, which is discussed in the next sections.

\begin{table}
\begin{tabular}{|l|l|l|l|l|}
\hline
GRB sample & N & $b \pm \delta_b$ & $\rho_{LT}$ &$P$\\\hline
ALL & 176& $-1.2\pm 0.1$ &-0.74& $4 \cdot 10^{-32}$\\
LONG-NO-SNe  & 128& $-1.0 \pm 0.1$ &-0.74 & $9 \cdot 10^{-24}$\\
LONG-SNe  & 19& $-1.5 \pm 0.3$ & -0.83 & $5 \cdot 10^{-6}$\\
LONG-SNe (A+B) & 7& $-1.9 \pm 0.3$ & -0.96& $3 \cdot 10^{-4}$\\
SE & 16& $-1.4 \pm 0.3$ & -0.71& $1 \cdot 10^{-6}$ \\
XRF & 25& $-1.6  \pm 0.3$ & -0.72 & $1  \cdot 10^{-6}$\\
\hline
\end{tabular}
\caption{Analyzed GRB samples. In the successive columns the table shows a GRB sample, a number N of events in the sample, the fitted correlation slope, $b$ and its error $\delta_b$, the Spearman correlation coefficient, $\rho_{LT}$, and the probability $P$.}\label{Table1}
\end{table}

The LONG-SNe analyzed in the present paper are listed in Table \ref{Table2}. There are cases of LONG-SNe that are excluded from our analysis either because they do not present plateau or because the paucity of the data do not allow a reliable fit with the W07 model.
 Within the LONG-SNe sample we applied a further classification, which is an update of the classification of Hjorth \& Bloom (2011). This classification identifies a `standard' sample of LONG-SNe with common properties, namely subsamples of the LONG-SNe based on the quality of the identification of SN associated with the GRB. The considered categories are as follows: A) strong spectroscopic evidence for a SN associated with the GRB; B) a clear light curve bump and some spectroscopic evidence suggesting the LONG-SN association; C) a clear bump on the light curve consistent with the LONG-SN association, but no spectroscopic evidence of the SN; D) a significant bump on the light curve, but the inferred SN properties are not fully consistent with other LONG-SN associations, the bump is not well sampled, or there is no spectroscopic redshift of the GRB; and E) a bump, either of low significance or inconsistent with other observed LONG-SN identifications, but with a spectroscopic redshift of the GRB. These five subcategories are indicated in the column category in Table \ref{Table2}.

\begin{table}
\begin{tabular}{|l|l|l|l|l|l|l|}
\hline
$GRB$ & $SN$ & $\log T^{*}_a$ &$\log L_a$&  $z$ & $Cat$ & Ref.\\ \hline
050416A & &2.99 & 46.76& 0.65 & D(X) & 29\\
050525A &05nc&2.89 &47.56& 0.61 & B & 30\\
050824 & &4.76 &45.25& 0.83 & E(X) & 31\\
051109B & &3.17 & 48.01& 0.08 & E & 45\\
060218 &06aj&5.14 &42.74& 0.03 & A(X) & 32,33,34\\
060729 & & 4.92& 45.97& 0.54 & E(X) & 35,36\\
070809 && 4.00& 44.22& 0.21 &E(X) & gcn 6732\\
080319B & &4.96 &45.60& 0.93 & C(X)& 39,40,41\\
081007 &08hw&4.21 & 45.86&  0.53 & B(X) & 42,54,44\\
090424 & & 2.86& 48.27 & 0.54 &E(X)& 53 \\
090618 & &3.18 &47.50& 0.54 & C(X)& 36,45\\
091127 &09nz&3.98 & 47.02& 0.49 & B(X) & 55\\
100418A & & 5.38&44.83& 0.62 & D(X) & 43\\
100621A & &3.52 &47.41 &0.54 & E& 66\\
101219B & &4.53 & 44.79& 0.55 & B(X) & 53\\
111228A & &3.80 &46.87& 0.71 & E & 46\\
120422A & &5.15 &43.53& 0.28 & A & 47\\
120729A & & 3.39& 47.04& 0.8 & D & 48 \\
130831A & &3.00 &47.14& 0.47 & B & 47\\
\hline
\end{tabular}
\caption{LONG-SNe divided into categories A,B,C,D, and E. We indicate with X the XRFs cases. Here the symbol of $\log=\log_{10}$. \newline 
References: 29) Soderberg et al. 2007; 30) Della Valle et al. 2006b; 31) Sollerman et al. 2007; 32) Pian et al. 2006; 33) Modjaz et al. 2006; 34) Sollerman et al. 2006; 35) Fynbo et al. 2009; 36) Cano et al. 2010; 39) Kann et al. 2008; 40) Bloom et al. 2009; 41) Tanvir et al. 2010; 42) Berger et al. 2008; 43); De Ugarte-Postigo 2010 private communication; 44) de Ugarte Postigo et al. 2011; 45) Perley et al. 2006; 46) Bersier et al. 2012; 47) Cano et al. 2014; 48) Cobb et al. 2011; 53) Sparre et al. 2011; 54) Jin et al., 2013. 55) Troja et al. 2012}
\label{Table2}
\end{table}

Our Swift LONG-SNe sample has a value of $\rho_{LONG-SNe}=-0.83$, which is higher than that of the LONG-NO-SNe sample, i.e., $\rho_{LONG-NO-SNe}=-0.76$. The A and B categories together (seven GRBs) present $\rho_{LONG-SNe_{A+B}}=-0.96$, the highest correlation with a probability $P=3.0 \times 10^{-4}$, thus confirming the existence of this tight correlation. We decided to gather only these two categories together because they have a stronger association with the SNe because of the presence of the spectral features. In fact, from category C to E there is no spectral information related to the association with the SNe. Therefore, this result may lead to the conclusion that the best-correlated sample has a clear spectroscopical identification of the underlying supernovae. 
 
We have observations by the previous mission of GRBs firmly spectroscopically associated with SNe, such as GRB $030329$ and GRB $031203,$ as well as GRB $980425$. However, these GRBs are not included in our sample because GRB $031203$ has only four data points in the light curve, which prevents us from any fitting; GRB $030329$ has no evidence of plateau, while GRB $980425$ has an indication of plateau (see diamond symbol in upper panel of Fig. \ref{fig3}), but the errors on the fitting parameters are not accurate enough, so this GRB is not included in our analysis either. We recall here that the last data point of GRB $980425$ belongs to the canonical X-Ray afterglow, even if it is at late time; for a reference to these non-Swift GRB light curves, see Fan et al. (2011) and Zhang et al. (2012). 

The high $\rho_{LT}$ for the LONG-SNe sample shows how on the basis of only obeying the Dainotti relation with a different slope from the LONG-NO-SNe, without any further selection criterion, we are able to select a homogeneous and observationally motivated subsample of GRBs. Indeed, before arriving at the conclusion that the LONG-SNe sample has the highest correlation coefficient, we tried several classifications based on the morphological structure of the light curves: for example, the $\chi^2$ of the fitted plateau or the flatness of the plateau itself. We conclude that the LONG-SNe subsample seems to be a better choice in terms of the highest correlation coefficient than any other selected on the basis of the morphology, spectral features of the light curves, and on the fitting parameters. Moreover, owing to the existence of the prompt-afterglow correlations \citep{Dainotti2011b,Dainotti2015b} and their theoretical interpretation \citep{Hascoet2014}, we also checked whether the Dainotti relation is tighter for certain correspondent values of the prompt emission parameters, such as $E_{peak}$, without finding any particular clustering of the Dainotti distribution. 
We also performed the same check for the spectral indices, $\beta_a$,  showing that there is not a significant trend with Dainotti distribution with $\beta_a$. This confirms previous results \citep{Dainotti2010}, but with a much wider sample, almost three times larger.

Also, Table \ref{Table1} shows that the slope determined for the (A+B) sample does not agree within $2.8$$\sigma$ with the LONG-NO-SNe sample; it only agrees within 3$\sigma$. To check whether this result is statistically significant, we applied the T-student test to the slopes of the two distributions, finding a probability $P=0.005$. This result is significant from a statistical point of view and it shows that there is a hint that the LONG-SNe sample may have a steeper coefficient than the $L_X-T^{*}_a$ relation for long GRBs. A slope of $-1$ implies a constant total energy. This result would imply that the standard energy reservoir that powers the plateau for the LONG-SNe is not constant. This condition is instead valid for the sample of LONG-NO-SNe. 

This evidence might possibly lead to a different theoretical interpretation for the plateau phase of the LONG-SNe sample. However, one can argue that, for
the LONG-NO-SNe sample, this difference may be due to selection effects from redshift evolution because the redshift range of this sample is much greater then the redshift range in which LONG-SNe are observed.
To investigate whether this difference is due to redshift evolution, we employed a robust statistical technique, the Efron and Petrosian (1992) method, to demonstrate the intrinsic nature of the Dainotti relation; see next section. In addition to this, another issue could be related to the fact that the steeper slope of the correlation is due to the presence of llGRBs in the sample. In  \S \ref{the beaming}, we show that the difference in the two slopes only remains significant statistically at the $10\%$ level when we correct the luminosity for the jet opening angle. We cannot confirm with the  present data if this difference in significance is due to the fact that we do not precisely know the angle.

\section{Motivation for the application of the Efron and Petrosian 1992 method}\label{EP method}

 In the current analysis we have nearly a 3$\sigma$ difference between the slopes of the LONG-SNe (A+B) sample and the entire LONG-NO-SNe sample. Because of the paucity of the LONG-SNe the Efron and Petrosian, EP, method (1992) cannot be accurately applied for this subsample of LONG-SNe. As mentioned before, this difference in the slopes can be due to redshift evolution, so to ground our results with a solid statistical analysis we quantitatively evaluate the difference between the intrinsic slope for LONG-NO-SNe and the observed slope for long GRBs through the procedure of the Efron \& Petrosian (1992) test. Then, if there is no difference between the observed and intrinsic slopes of the LONG-NO-SNe, we infer that there also should be no difference between the observed and intrinsic slopes for the LONG-SNe sample and, therefore, the comparison between the two observed slope (A+B) categories and the LONG-NO-SNe (128 GRBs) is appropriate. In addition, we stress that the LONG-SNe are placed at very small redshift, so the effect of redshift evolution is negligible in any case.
The details of the method are shown in the Appendix. We demonstrated that the intrinsic slope for the LONG-NO-SNe obtained through the EP method is $b_{int}=-1.02 \pm 0.12$. Therefore, the agreement is the same between this measurement and that of the observed slope of the LONG-NO-SNe sample presented in Table \ref{Table1}. Therefore, based on this method we confirm the comparison with the observed sample is valid and that the two slopes are statistically different. 

\section{The Dainotti relation corrected by the beaming angle}\label{the beaming}

The jet opening angle of GRBs is a relevant feature for determining the nature of the progenitor, and for shedding light into the relativistic outflow and the total energy of the burst. Unfortunately, a reliable determination of the jet opening angle requires a broadband measurement of the GRB afterglow from X-ray to radio observations and from minutes to days after the prompt gamma ray emission, which can be very challenging to detect. Thus, very few of all detected GRBs have measured jet angles. 

For this  reason, in the current paper and previous papers we assume that the emission is isotropic during the plateau phase. Even if this assumption gives a rough estimate of the luminosity, more precisely an overestimation of its true value, it is a reasonable assumption. Otherwise the plateau luminosity should be computed considering a jet opening angle, which can be in principle different for each GRB. 

\begin{figure}[!tbh]
\includegraphics[width=9cm]{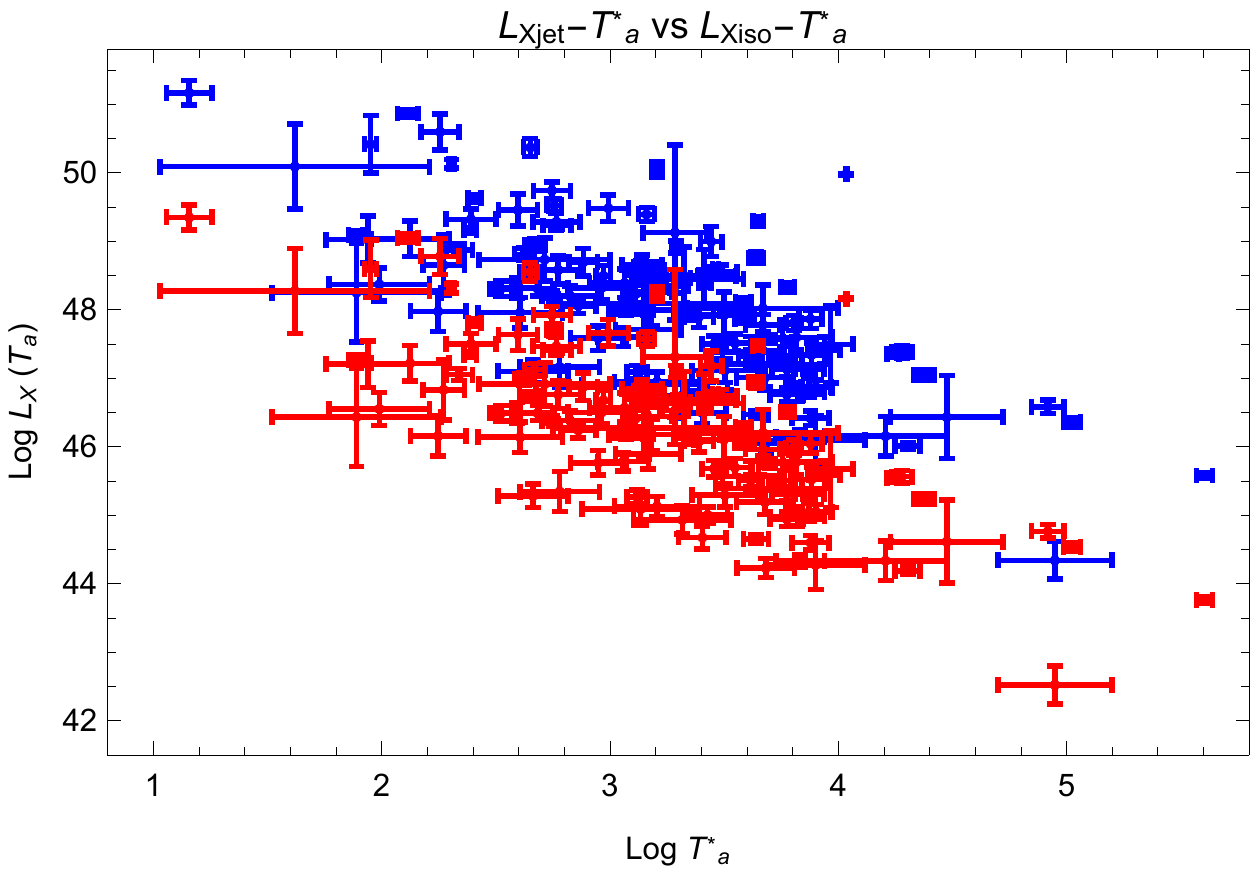}
\includegraphics[width=9cm]{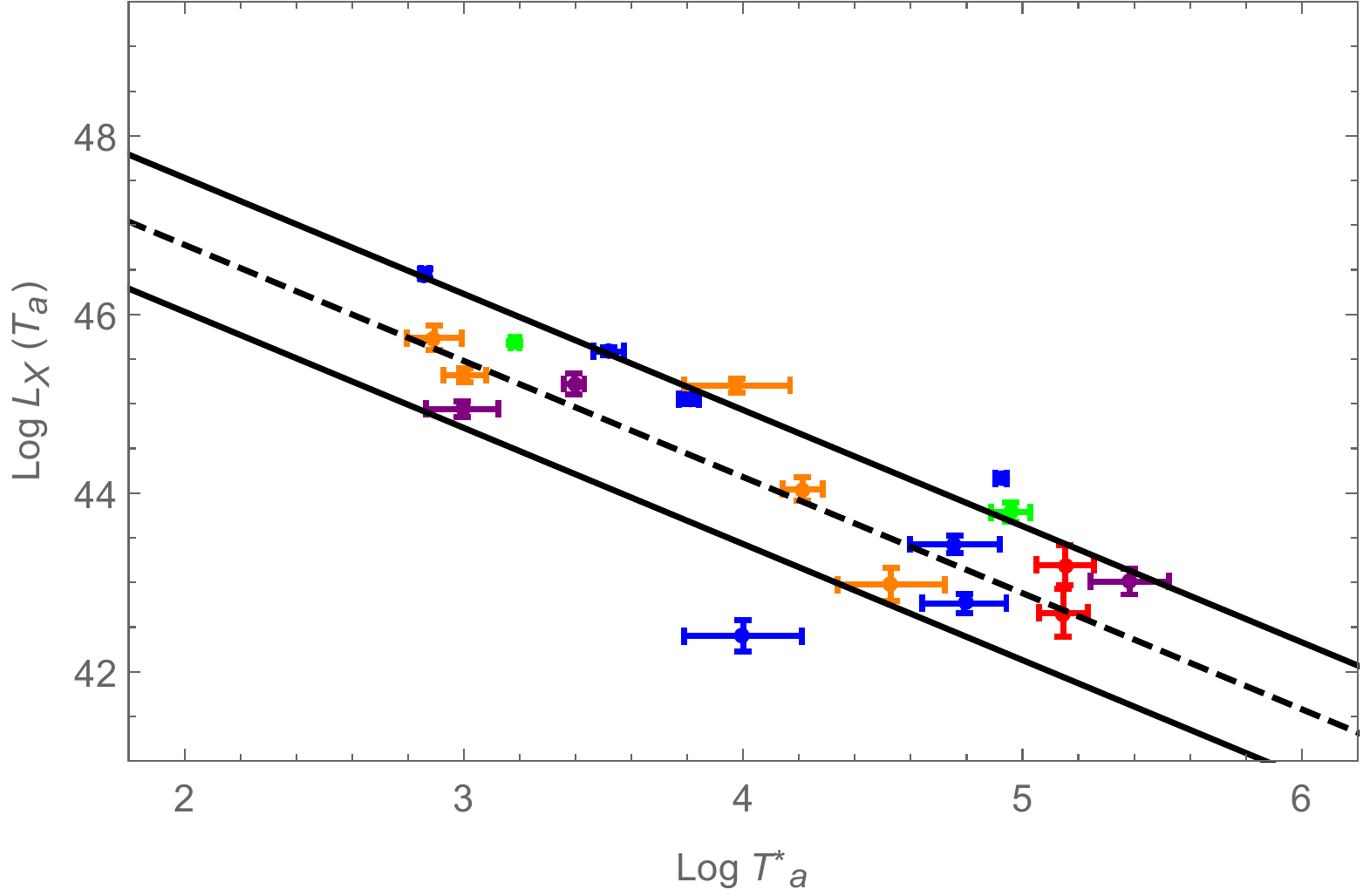}
\caption{Upper panel: The analyzed $\log L_X$ vs. $\log T_a^*$ distributions for LONG-NO-SNe computed assuming isotropic luminosity (blue points) and beamed luminosity (red points). Lower panel: The analyzed distribution of  $\log L_{X,jet}$ vs. $\log T_a^*$ corrected for beamed luminosity for the GRB-SNe  divided in categories as explained in Table 2. Category A: red points; B: orange; C: green; D: purple; and E: blue.}
\label{fig2bis}
\end{figure}

In this section we assume that during the plateau phase the emission in a regular GRB is still beamed to a GRB jet opening angle $\theta_j \approx 10^{\degree}$. We assume $\theta_j$ to be the same for all regular or high luminosity GRBs and that the GRB is seen on axis. Thus, the isotropic equivalent luminosity, $L_X$, overestimates the true luminosity by a factor of $\approx \theta_j^{2}/4$. Also, a scatter in $\theta_j$ introduces an additional scatter in $L_X$, even if all GRBs had the same total energy. Therefore, we add a scatter of $10\%$, namely $0.1 \degree$ to each GRB angle. We checked that within the distribution of the LONG-NO-SNe sample there is no event with peak luminosity $L_{iso}\leq 10^{48}$ $erg s^{-1}$, thus none of the GRBs of this category are llGRBs, and we can thereby correct their luminosities with the same jet opening angle for simplicity. As has been pointed out by Liang et al. (2007), the llGRBs have different beaming angles,  which are typically wider than $31^{\degree}$.
We show in Fig. \ref{fig2bis} the difference between an isotropic distribution (blue points) and a distribution of beamed luminosities $L_j=(1-cos\theta)*L_j$ (red points) for the 128 LONG-NO-SNe sample. The slope of the distribution remains the same, while the intercept is larger for the isotropic luminosities. Thus, the results quoted in Table $1$ remain the same for the LONG-NO-SNe sample even with the beamed luminosities. 

A low luminosity GRB has a much wider $\theta_j$ and a factor on the order of a few at most, thus its emission is not beamed or is beamed to a very small degree (see, e.g., Soderberg 2006 Nature; Liang et al. 2007). In such a case $L_X$ is much closer to the true isotropic X-ray luminosity.
Therefore, we plot $L_X$ of llGRBs computing their beaming angles. In the case of GRB 060218, we chose $1.4$ radians equivalent to $80^{\degree}$ according to Soderberg et al. (2006). For the GRB 120422A, we quote the parameters presented in Schultze et al. (2014) and we use the time of the break presented there to convert it to a jet opening angle according to Sari et al. (1999) and Frail et al. (2001) :

\begin{equation}
j = 0.057 t_{j,d}^{3/8} ((1 + z)/2)^{3/8}E_{\gamma,iso,53}^{1/8} (\eta/0.2)^{1/8}*(n_0/0.1)^{1/8} deg
\end{equation}

where $t_{j,d}$ is in days, $E_{iso,53}$ is in units of $10^{53}$ erg, and $n_0$ is in units of $ cm^{3}$. If we assume $n_0=0.1$ and $\eta=0.2,$ we obtain $\theta_j \approx 18^{\degree}$.
For GRB 051109B, we do not have an explicit measure of the angle, however, this burst has been classified as ll-GRB (Bromberg et al. 2011). It has been established that llGRBs have a jet angle typically wider than $31^{\degree}$ \citep{Liang2007}. Thus, we associate GRB 051109B with an opening angle of $32^{\degree}$. In Fig. 3, we show the distribution of the GRB-SNe, correcting these three GRBs for the mentioned jet opening angles and correcting the other regular GRBs for the same jet opening angle of $10^{\degree}$. The absolute value of the slope for the 19 GRB-SNe is much smaller, $b=-1.23 \pm 0.24$, while the scatter is slightly smaller than the isotropic luminosity computation and $\rho=-0.78$. Correcting the isotropic luminosity with the introduction of these jet opening angles confirms the claim that only the E category sample are outliers of the correlations, thus showing more clearly that the most correlated sample is possibly driven by its more secure association with the SNe; see next section for additional details. 
Also the correlation coefficient of the GRB-SNe (A+B) category is pretty high, $\rho=-0.86$, and the slope is $b=-1.35 \pm 0.24$. The slopes are flatter than those in Table 2, but again the error bars are slightly smaller. We compute the T student to check if the difference between the slopes of LONG-NO-SNe (128 GRBs) and the LONG-SNe (A+B, 7 GRBs) is statistically significant. We found that $P=0.10$. Since our hypothesis confirmed only at $10\%$ level when we correct the luminosity for the beaming angle, this result opens the debate that the difference between the SNe-LONG and LONG-NO-SNe could be due to the isotropic approximation. However, since the precise determination of the angle is not known, we cannot achieve a definite conclusion with the current available data.
 
\section{LONG-SNe: A well-correlated sample}\label{standardizable candle}
Previously we determined statistically that the difference between the slopes is significant and does not result from selection bias due to redshift evolution, even though the presence of llGRBs in the LONG-SNe sample raises the probability that the slopes are the same to within $10\%$. Now we discuss the Dainotti correlation outliers for the LONG-SNe and the connection of this relation to the SNe properties associated with the A+B sample. 
As presented in Table \ref{Table1} correlation coefficients of the luminosity-time distributions increase when one moves from the LONG-NO-SNe sample to the LONG-SNe one and, finally, to its subsample of the LONG-SNe (A+B) seven events, where the correlation is very high with $\rho=-0.96$. There are two clear outliers in the LONG-SNe Dainotti relation GRBs (GRB 070809 and GRB 060729); see blue points outside the 1$\sigma$ range of the best-fit line in the right panel of Fig. \ref{fig2}. Those two GRBs belong to the E category. Thus, this may mean that indeed these GRBs are not associated with SNe and they instead belong to the LONG-NO-SNe sample.

One question is whether the LONG-SNe is a feature exclusively owned by the distinct class showing the higher degree of correlation among $L_X-T^{*}_a$ and a steeper slope. If so, then the association with SNe may be the main reason for the high correlation coefficient for the Dainotti relation and for its different slope for the LONG-NO-SNe.  
In testing this hypothesis, one may note that there are two GRBs at relatively low-z for which no bright associated SN is observed because the upper limit to the luminosity of the SN that is possibly associated with it was at least two orders of magnitude fainter \citep{dellavalle2006,fynbo2006,gal-yam2006,Fruchter2006} than the peak luminosity of broad-lined SNe-
Ibc normally associated with GRBs. 

Here, GRB 060505 is slightly off the 1$\sigma$ correlation line (dark yellow point in Fig. \ref{fig2}), while GRB 060614 is within $1\sigma$ in the Dainotti relation. This burst is not represented in the Fig. \ref{fig2}, because as we have discussed previously GRB 060614 is also classified as a short GRB with an extended soft emission \cite{Zhang2007}. 
Very recently Yang et al. (2015) discovered a near-infrared bump that is significantly above the regular decaying afterglow. This red bump is inconsistent with even the weakest known supernova. It can arise from a Li-Paczy\'{n}ski macronova, the radioactive decay of debris following a compact binary merger. If this interpretation is correct, GRB 060614 arose from a compact binary merger rather than from the death of a massive star and, for these reasons, it can also be identified within the SE category. Therefore, this particular GRB does not constitute evidence against our hypothesis for its peculiar nature. The present statistics of events available for the analysis is very limited, but GRBs showing a clear association with the underlying SN form that is in some way a physically distinct sample with a tight Dainotti relation with a different slope.

Moreover, it has been suggested that optical light curves of LONG-SNe under certain conditions follow a tight relation between the peak luminosity and the light curve decay timescale (`stretch'), which is similar to what has been found for SNe Ia \citep{cano2014,cano2015,cano2016}. Light curves of all LONG-SNe in categories A and B, except for GRB 081007  associated with SN 2008hw, are included in the sample from which the luminosity-stretch relation has been derived \cite{cano2014}.

Indeed, GRB 081007 is a possible outlier in this relation: Jin et al. (2013) suggested that it is fainter than SN 1998bw by a factor of $2$ without considering any stretch, which is off the peak-stretch relation fit line. According to our analysis, the stretch parameter cannot be much shorter than unity even in the case of a relatively faint peak luminosity, thus placing this LONG-SN as a possible (mildly) outlier in the SN property. On the other hand, it is not an outlier in the Dainotti relation. However, given a single event that does not follow the `standard' relation in the SN properties (i.e., GRB 081007 in our sample) this does not contradict our hypothesis that the Dainotti relation for the A+B sample and the peak luminosity stretch relation could possibly have the same origin.

To further investigate whether the LONG-SN association, which shows a strong $L-T$ relation and a steeper slope than the LONG-NO-SNe, (and the SN property), is a distinct feature in the subsample of long GRBs, it is important to increase the sample of GRBs for which both deep optical light curve at an expected SN-bump phase and good early-phase XRT light curve are available. We will be able to achieve this task when we are able to detect LONG-SNe at higher redshift than they are currently observed. This may be feasible with the very large, $30-40$ m telescopes, such as the European Extremely Large Telescope (E-ELT), the Thirty Meter Telescope (TMT), and the Giant Magellan Telescope (GMT).

\section{Summary and conclusions}\label{Conclusions} 
Our analysis of the GRBs-SNe compared to the LONG-NO-SNe leads to several interesting findings. First, after categorizing and dividing the whole sample into subsamples, we discovered that the LONG-SNe have a higher Spearman correlation coefficient, $\rho$, between the luminosity and rest-frame duration of the
plateau, than any other analyzed subsamples. Moreover, the cases with most firm spectroscopic associations (categories A and B) form a sample of GRBs with a highly correlated $L_X-T^{*}_a$ relation, reaching an almost perfect anti-correlation $\rho=-0.96$ with a probability $P=3.0*10^{-4}$, but with a much steeper slope than the LONG-NO-SNe sample. The difference between the two slopes is significant at the $P=0.005$ level. This result possibly leads to a new scenario that the LONG-SNe sample is a different population from the LONG-NO-SNe. Thus, in the future we should be able to see more cases of real long GRBs-NO-SNe, such as GRB $060505$.
Indeed, we have shown by the means of a robust statistical test that this difference is not an artifact of a steepening of the slope from GRB selection bias  and, therefore, may possibly indicate that for the LONG-SNe the energy reservoir of the plateau does not remain constant unlike for the LONG-NO-SNe. This may open new perspectives in future theoretical investigations of the GRBs with plateau emission and associated with SNe.
We need to place a caveat on this statement since we only have seven GRBs in the current LONG-SNe (A+B) sample and, additionally, the difference between the SNe-LONG (A+B) and LONG-NO-SNe sample is only statistically significant at the $10\%$ level when we consider the beaming correction. Thus, one can argue that the difference in slopes can be partially due to the effect of the presence of llGRBs in the LONG-SNe sample that are not corrected for beaming. However, the beaming corrections are not very accurate due to indeterminate jet opening angles, so the debate about this difference remains open and it can only be resolved when we have more data.

In addition, all LONG-SNe associated with spectroscopic evidence, with exception of GRB 081007/SN 2008hw, also obey the peak-magnitude stretch relation, similar to that which characterizes the SNe Ia standard candles. This may suggest that the same physical mechanism could be responsible for both the Dainotti and Philipp relations. 

Because of the indication that the LONG-SNe sample may constitute a physical motivated subsample, it is advisable to separate it from the LONG-NO-SNe sample if in future the LONG-SNe sample can be used as a candle that can be standardized to measure cosmological distances and constrain cosmological parameters. \newline

\section{Acknowledgments}
This work made use of data supplied by the UK Swift Science Data Centre at the University of Leicester. We thank A. Mizuta, J. Matsumoto, and H. Ito for their critical reading of the manuscript. We are particularly grateful to M. Barkov and M. Ostrowski for their discussions about the comparison between Long-NO-SNe GRBs and LONG-SNe sample and valuable comments on the manuscript. We are very thankful to V. Petrosian for his comments on the Efron \& Petrosian method and for his remarks on the manuscript. We are very grateful to A. Boria for her initial contribution to the fitting of part of the data sample during her summer internship at RIKEN and J. Arratia for the NSF financial support for her internship. M.D is grateful for the initial support from the l'Oreal Fellowship and the JSPS Foundation (No. 25.03786), and the fruitful discussion at the ITHES Group in RIKEN. M.G.D is grateful to the Marie Curie Program because the research leading to these results has received funding from the European Union Seventh FrameWork Program (FP7-2007/2013) under grant agreement N 626267.  S. N. is grateful to JSPS (No.24.02022, No.25.03018, No.25610056, No.26287056) $\&$ MEXT(No.26105521). K.M. acknowledges financial support by JSPS Grant-in-Aid for Scientific Research (No. 23740141 and 26800100). The work by K.M. is partly supported by WPI Initiative, MEXT, Japan.

\begin{appendix}\label{the EP method}
\section{Description of the Efron \& Petrosian (1992) method}
 
\begin{figure}
\includegraphics[width=9cm]{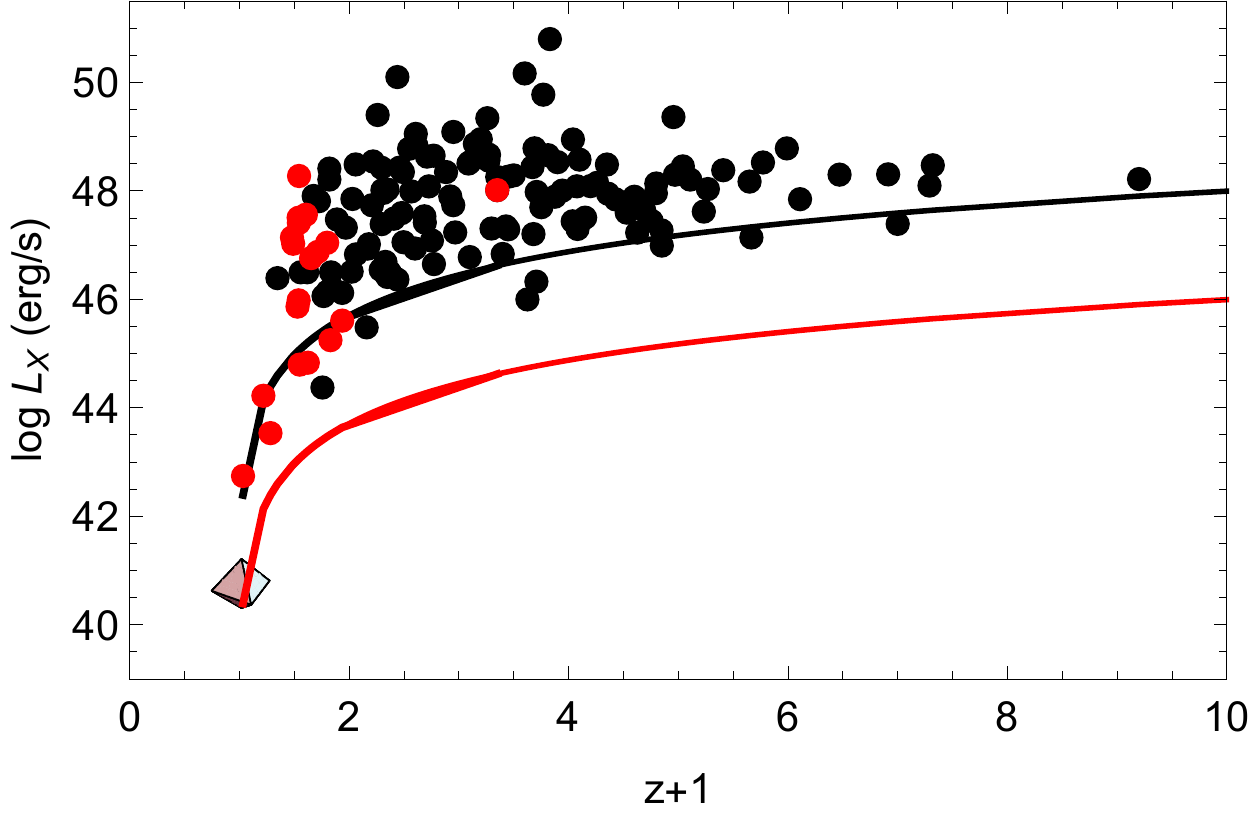}
\includegraphics[width=9cm]{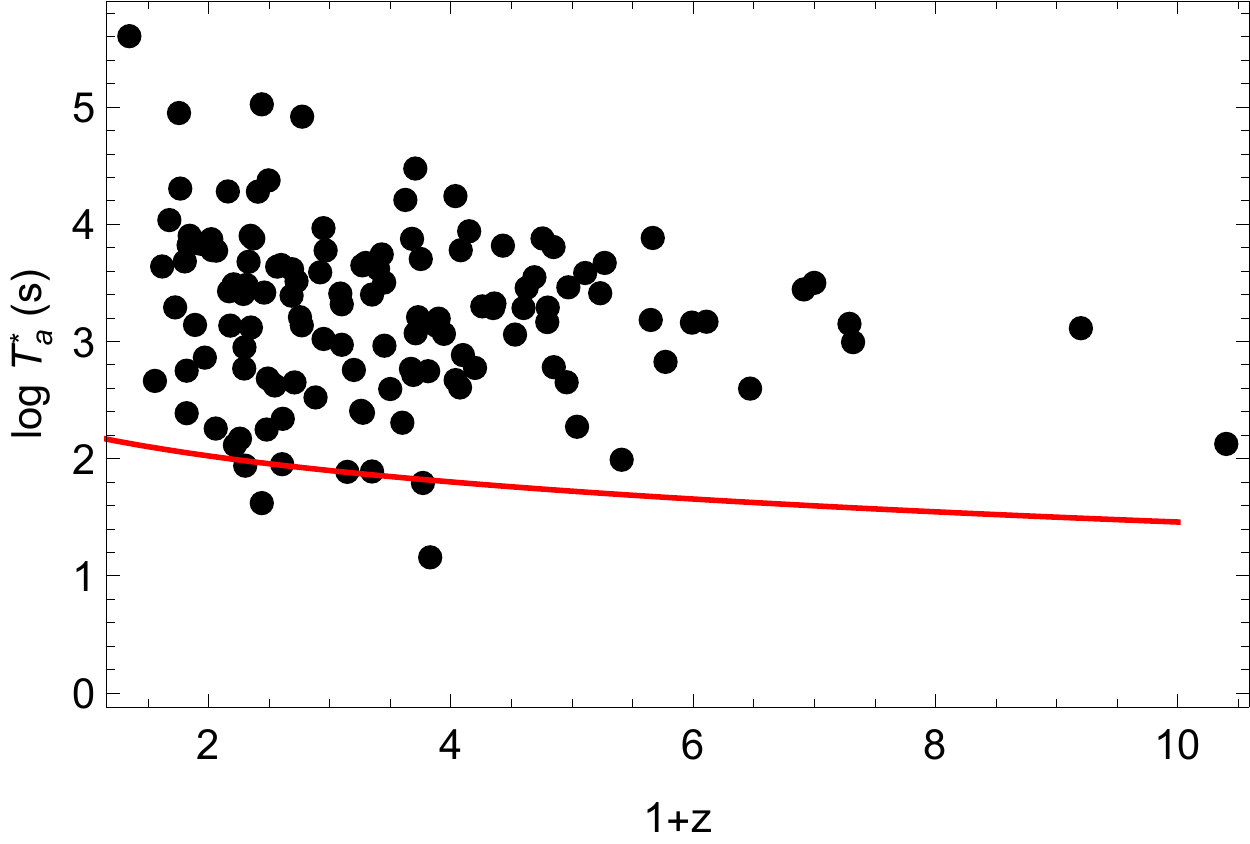}
\caption{Upper panel: Distribution of $\log L_X$ and redshift of the LONG-NO-SNe. The limiting luminosity, obtained using the flux limit such as $2.0 \times 10^{-12}$ erg cm$^{-2}$, shown with a solid black line, better represents the limiting luminosity of the sample. In the same panel, we show the distribution of the LONG-SNe sample in red. The solid red line represents the XRT limiting luminosity based on the XRT sensitivity in $10^{4}$s observation, facilitating the observation of GRB 980425.
{\bf Lower panel}: Distribution of the rest-frame time $\log T^*_{a}$ and the redshift, where we denote with the red solid line the limiting rest frame time, $\log T^{*}_{a,{\rm lim}}=\log T_{a,{\rm lim}}/(1+z)$, with a red solid line.}
\label{fig3}
\end{figure}

The Efron \& Petrosian technique, applied for GRBs \citep{Petrosian2009,Lloyd1999,Lloyd2000}, allows us to compute the intrinsic slope of the correlation by creating new bias-free observables, called local variables and denoted with the symbol ${'}$, for which the redshift evolution and selection effects from instrumental thresholds are removed.  For redshift evolution we denote the dependence of variables on the redshift. We name luminosity and time evolution the dependence of the plateau luminosity and plateau duration on the redshift, respectively. For details about the formulation of the method, see Dainotti et al. (2013a, 2015b).
First, we need to determine the instrumental threshold limits, such as the flux limit and time limit to remove the selection effects. The limiting flux, $F_{lim}$, is defined as the minimum observed flux at the end of the plateau for a given redshift for which the minimum luminosity is given by $L_{lim}=  4 \pi D_L^2(z) \, F_{lim} K$. The value $F_{lim}$ is chosen to be $1.0 \times $10$^{-12}$ erg cm$^{-2}$, so that we have $121$ GRBs in the sample and $L_{min}(z)$ is presented with a black solid line in the upper panel of Fig. \ref{fig3}. This limit is a good compromise between keeping a large sample size and being representative of the sample itself. The red solid line instead represents the
limiting luminosity for the LONG-SNe sample. We also represented GRB 980425 with a diamond
symbol to show that its plateau could have been observed within the limiting flux set by Swift. With the same criterion of choosing a large sample size and whether this limit should be representative of the sample, we also determined the observed minimum time for the plateau, $T^{*}_{a,{\rm lim}}= 309/(1+z)$ s, shown by a red solid line in the lower panel of Fig. \ref{fig3}. This choice allows the inclusion of $121$ GRBs.

With the modified version of the Kendall tau statistic, we obtain functional forms that are the best descriptors of the luminosity and time evolution for the LONG-NO-SNe sample. 

\subsection{The luminosity and time evolutions}

To determine the local variables, we need to identify the functional forms, $g(z)$ and $f(z)$, which define the evolution of $L_{a}$ and $T^{*}_{a}$, respectively. Hence, the redshift-independent variable is $L'_{a} \equiv L_{a}/g(z)$ and $T'_{a} \equiv T^*_{a}/f(z)$. In this way dividing each variable for the appropriate evolutionary function, the new variables, thus, are not correlated with the redshift anymore. In this way, we overcame the problem of the selection bias on the redshift.
These evolutionary functions are parametrized by simple functions, such as 
\begin{equation}
g(z)=(1+z)^{k_{L_{a}}}, f(z)=(1+z)^{k_{T^{*},a}}
\label{lxev}
\end{equation}
 
More complex evolution functions lead to comparable results; see Dainotti et al. (2013a, 2015b).

\begin{figure}
\includegraphics[width=9cm]{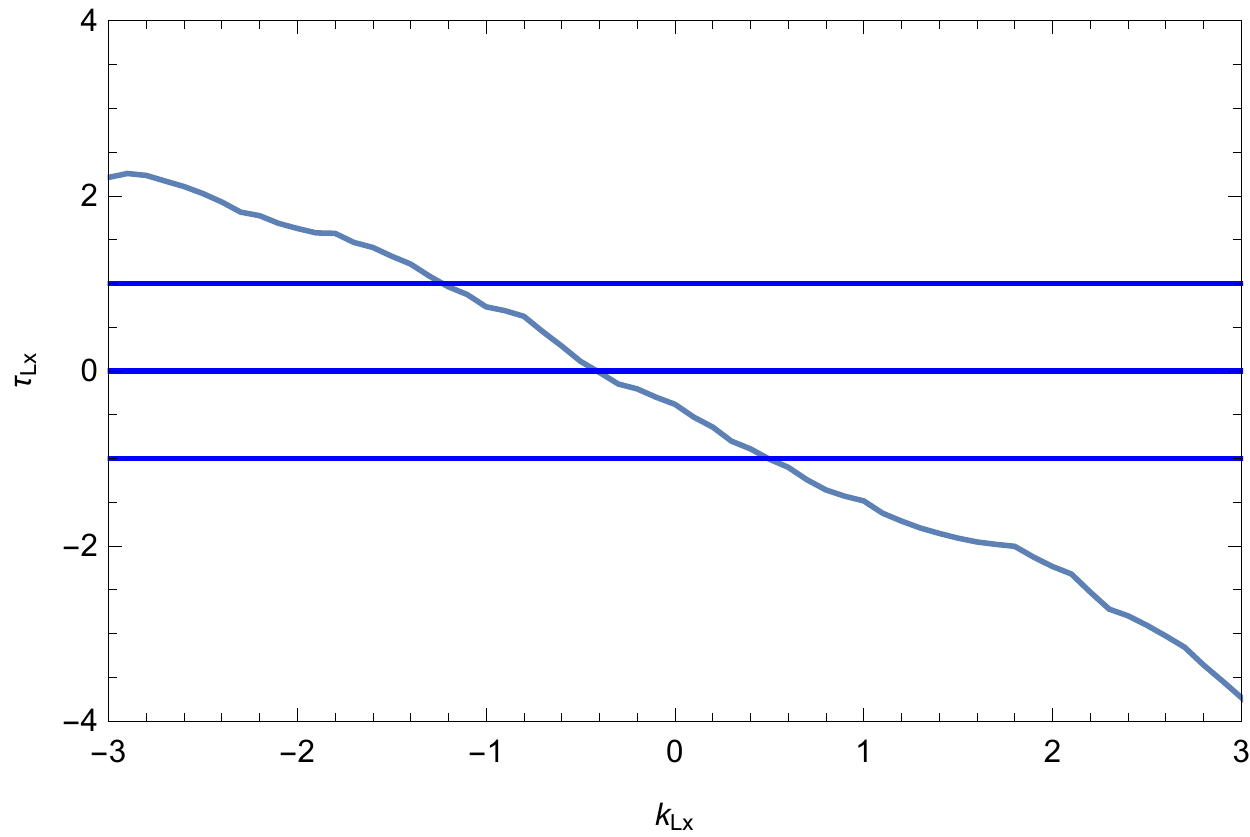}
\includegraphics[width=9cm]{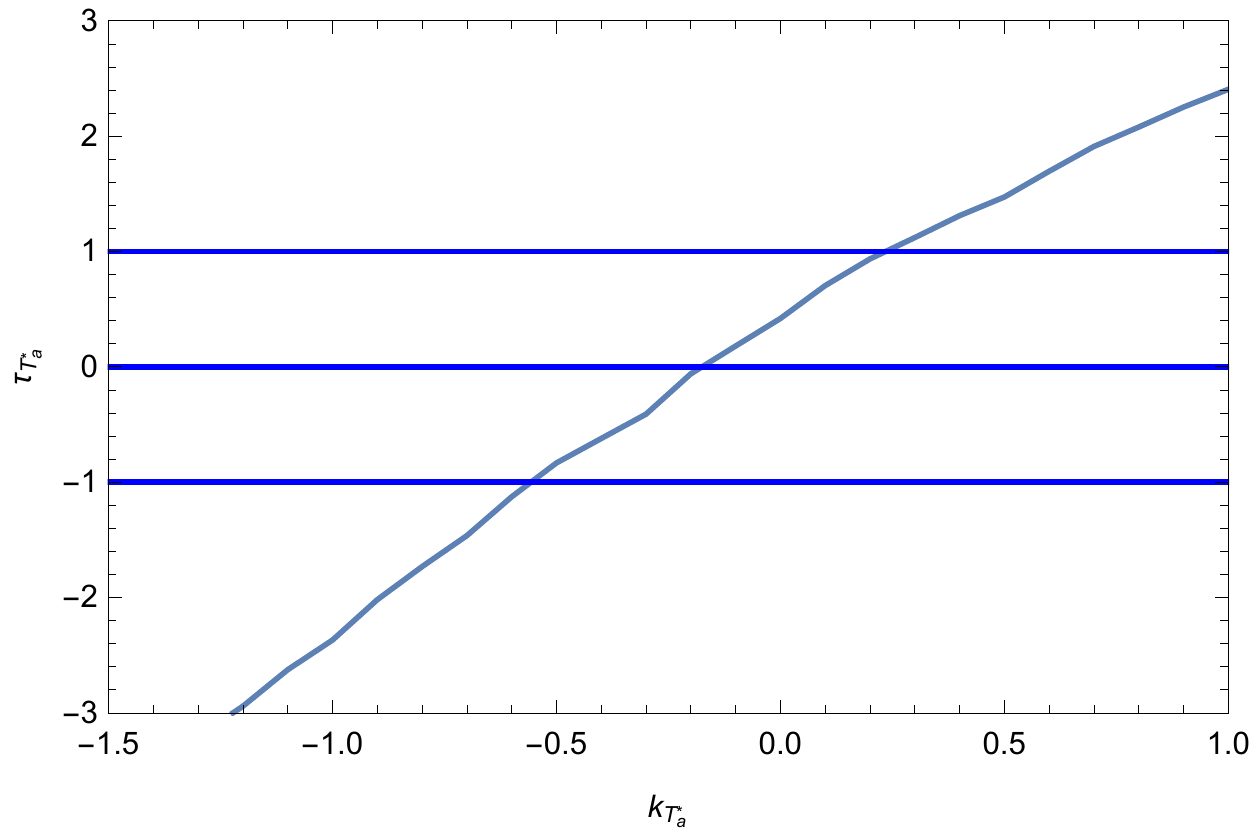}
\caption{Upper panel: Test statistic $\tau$ vs. $k_{L_{a}}$, the luminosity evolution defined by Eq. \ref{lxev} using a simple power law as $g(z)$. Lower panel: Test statistic $\tau$ vs. $k_{T^{*}_{a}}$, the time evolution defined by Eq. \ref{lxev}}
\label{Fig4}
\end{figure}

With the specialized version of Kendell's $\tau$ statistic, the values of $k_{L_{a}}$ and $k_{T^{*}_{a}}$ for which $\tau_{L_{a}} = 0$ and $\tau_{T^{*}_a} = 0$ are the best fit to the luminosity and plateau time evolution, respectively, with the 1$\sigma$ range of uncertainty given by $| \tau_x | \leq 1$. Plots of $\tau_{L_{a}}$ and $\tau_{T^{*}_{a}}$ versus $k_{L_{a}}$ and $\tau_{T^{*}_{a}}$ are shown in the upper and lower panel of Fig. \ref{Fig4}, respectively. With the determination of $k_{L_{a}}$ and $k_{T^{*}_a}$ we are able to compute accurately the local observables $T{'}_{a}$ and $L{'}_{a}$. 

There is a low luminosity and time evolution in the afterglow, $k_{L_{a}}=-0.40_{-0.83}^{+0.89}$ and $k_{T^{*}_{a}}=-0.17_{-0.37}^{+0.41}$ for the simple power law functions. To make our comparison between the observed and intrinsic slopes even more solid than that presented in Dainotti et al. (2013a), we also include the errors on the luminosity and time evolution in the evaluation of the intrinsic slope so that we can apply the D'Agostini 2005 method, which takes into account both errors on the variables. In this way the same D'Agostini method has been applied for both the intrinsic and observed slopes making this comparison more reliable. The resulting intrinsic slope is $b_{int}=-1.02 \pm 0.12$. Therefore, we note that there is the same agreement between this measurement and that with the observed slope of the LONG-NO-SNe sample presented in Table \ref{Table1}. Therefore, based on this method we demonstrated the comparison between the two observed samples was performed appropriately. 

\section{GRBs associated with SNe and their luminosity nature} \label{LONG-SNe and prompt emission}

\begin{figure}
\includegraphics[width=9cm]{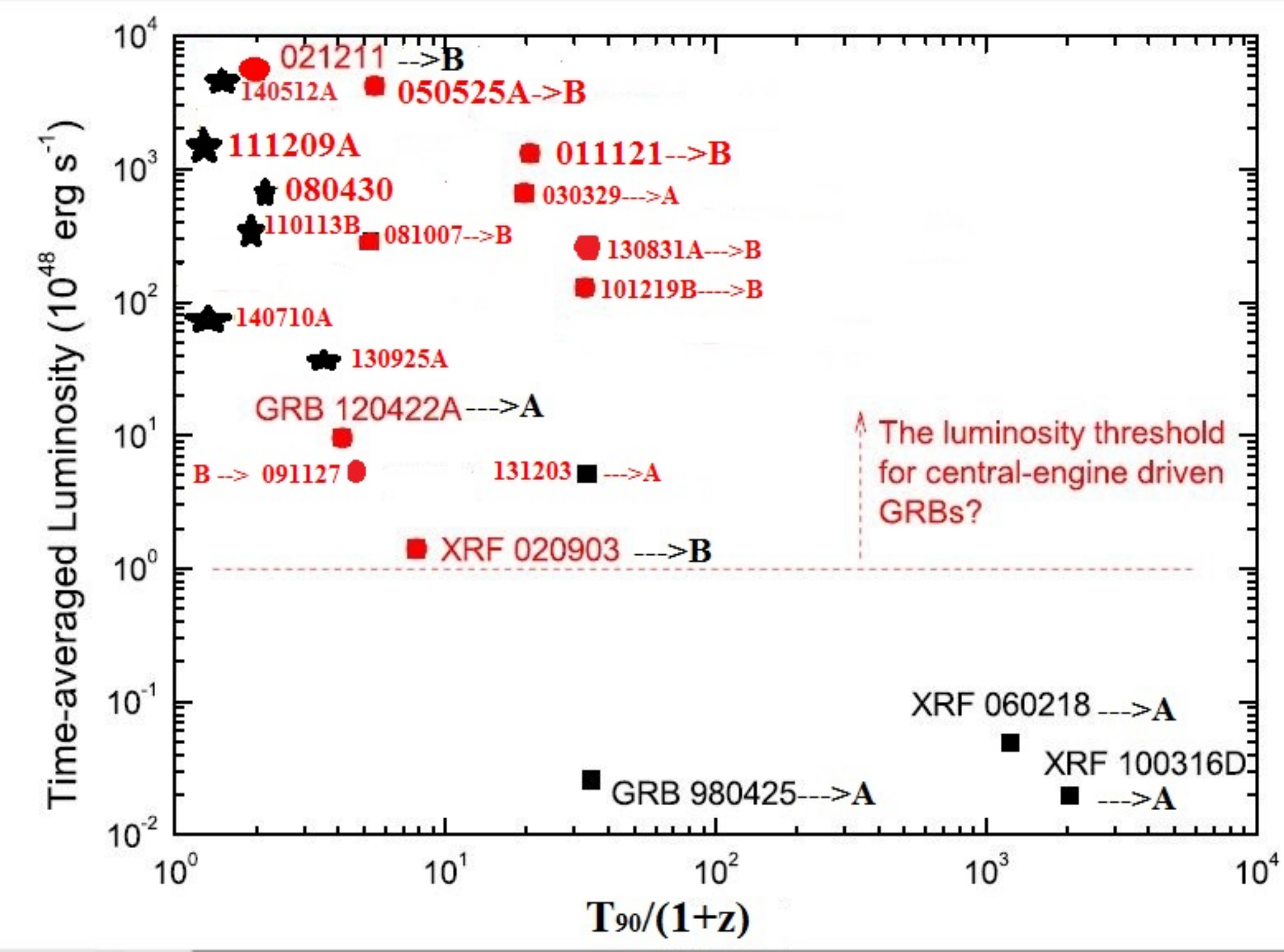}
\caption{Supernova-associating GRBs in the time-prompt, emission averaged luminosity - $T_{90}/(1 + z)$ plane. The red symbols denote engine-driven
GRBs, while the black symbols denote the possible shock breakout GRBs suggested in some of the literature. The red dashed line gives a rough
threshold above which a successful jet is possible. The A and B letters refer to the classification of Hjorth \& Bloom (2001). The star symbols indicate instead the LONG-NO-SNe at $z<0.79$.
\label{fig2a}}
\end{figure}

In order to better understand the properties of the A+B category, we also investigate the prompt emission properties of the LONG-SNe sample and the relation with their luminosity nature. To this end, we divide the GRBs in our A+B sample in low luminosity, intermediate and normal luminosity GRBs according to the literature classification. 

Our goal is to try to understand if there is an underlying mechanism that drives GRBs with plateau in the A+B category according to their observational properties. 
The relation between long GRBs, with and without the SNe associated (LGRBs), and llGRBs is a long-standing problem because, on the one hand, their high-energy emission properties are substantially different, implying a diverse gamma ray source, and, on the other hand, they both are associated with SNe with broad-line Ic, pointing at a similar progenitor and explosion mechanisms.

The LGRBs are luminous ($10^{50}-10^{52}$ erg $s^{-1}$), hard  ($\ge 100$ keV), highly variable, and narrowly collimated with a typical duration of $10-100$ s \citep{Piran2004}, while llGRBs are fainter by about four orders of magnitude ($10^{46}-10^{48}$ erg $s^{-1}$), relatively soft ($\approx 100$ KeV), not highly beamed, and show no significant temporal variability over their entire duration, which is often longer than $1000$ s \citep{Kulkarni1998,Soderberg2006,Kaneko2007}.

At low redshift, several GRBs have been discovered with peak luminosities ($L_{iso} \leq 10^{48.5}$ erg $s^{-1}$) that are much lower than the average of the more distant GRBs ($L_{iso} \geq 10^{49.5}$ erg $s^{-1}$). The properties of several llGRBs indicate that they are possibly generated by a breakout of a relativistic shock from the surrounding massive wind of the progenitor star \citep{Colgate1974,Tan2001}, as opposed to the emission from ultrarelativistic jets that originates LGRBs, also known as `cosmological' or `normal' GRBs.

In addition, theoretical calculations show that the gamma rays seen in llGRBs cannot be produced in the same environment where the gamma rays in LGRBs are generated \citep{Bromberg2011}. Notwithstanding these differences, it is highly debated how the populations of llGRBs and LGRBs are connected and whether there is a continuum between them. 

\subsection{The A+B category}
Starting from the A category, the case of GRB 060218, associated with SN2006aj, has been classified as a low luminous event \citep{Pian2006,Soderberg2006,Maeda2007,Zhang2012}. It is the second closest GRB associated with SNe, $z=0.033$ after the closest event, GRB980425 at $z=0.0085$, with an isotropic energy in the prompt emission $E_{iso}=4 \times 10^{49}$ erg and $T_{90}=2100 \pm 100$ s. It has been claimed that this event has a signature of being caused by a shock breakout in which the jet failed to pierce the stellar envelope.

GRB 120422A is one of the very few examples of intermediate luminosity GRBs with a $\gamma$-ray luminosity of $L_{iso}=10^{48.9}$ erg $s^{-1}$ that have been detected up to now 
\citep{Zhang2012,Schulze2014}. The interpretation of this burst points to a central engine origin in contrast to a shock breakout origin, employed to interpret some other nearby low luminosity supernova GRBs.
Comparing the properties of GRB 120422A and other supernova GRBs, Zhang et al. (2012) suggested that the main criterion to distinguish engine-driven GRBs from shock breakout GRBs is the time-averaged, prompt $\gamma$-ray luminosity. The isotropic peak luminosity of this GRB is $L_{peak} \approx 10^{49}$ erg $s^{-1}$, while the $E_{iso}=4.5 \times 10^{49}$ erg. 

More specifically, Zhang et al. (2012) compared GRB 120422A with GRB 060218. They used a similar categorization as that adopted by Hjorth \& Bloom (2011), which we also follow in the present paper. They pointed out that GRB 120422A has several peculiar features: the shortest $T_{90}$, high initial X-ray luminosity (e.g., greater than that of GRB 060218 by a factor of $100$), a steep temporal decay slope, and an X-ray afterglow plateau significantly brighter than GRB 060218 in the same time window (i.e., $10^{4}-10^{5}$ s). Nevertheless, the total prompt emission $\gamma$/X-ray energies of these bursts are comparable. This suggests that a much higher energy is carried by the relativistic outflow in GRB 120422A.

We now investigate the B category sample. 
 Zhang et al. (2012) showed the supernova-associating GRBs in the time-averaged prompt luminosity - $T_{90}/(1 + z)$ plane, which led to their claim that above $10^{48}$ erg $s^{-1}$, an engine-driven GRB is possible. Shock breakout luminosity cannot be much higher than this value. This consideration places GRB 060218 as a shock break out GRB, while GRB 120422A, 101219B, 081007, and 050525A as central engine driven GRBs. 

We here update the time-averaged prompt luminosity - $T_{90}/(1 + z)$ plane adding the missing GRBs in our (A+B) sample, namely GRB 091127 and 120831A.

GRB 091127 \citep{Troja2012} presents a standard afterglow behavior that is typical of cosmological long GRBs, however, its low-energy release ($E_{iso} \leq 3 \times 10^{49}$ erg), soft spectrum, and unusual spectral lag identifies this GRB as a subenergetic burst. Its $T_{90}=7.1 \pm 0.2$ (s) and its average prompt luminosity, $L_{iso}=6.29 \times 10^{48}$ erg $s^{-1}$, also places this GRB as an engine-driven mechanism event.

GRB 130831A is classified as a normal luminous GRBs, since $E_{iso}= 4.6 \pm 10^{51}$ erg in the $20$ keV -- $10$ MeV range \citep{Hagen2013,Barthelmy2013b} and its $T_{90}=32.5 \pm 2.5$ s, thus its average prompt luminosity $L_{iso}=E_{iso}/(T_{90}/(1+z))=2.08 \times 10^{50}$ erg $s^{-1}$ places this GRB in the time-averaged prompt luminosity - $T_{90}/(1 + z)$ plane as a driven engine GRB. 

To conclude, we plot an updated diagram of the supernova-associating GRBs in the time-averaged prompt emission luminosity -$T_{90}/(1 + z)$ plane (see Fig. \ref{fig2a}) in which we add the spectroscopic category of the SNe associated with GRBs (A+B) and, also, the LONG-NO-SNe at small redshift, $z \leq 0.79$, for details on the comparison between LONG-NO-SNe and LONG-SNe; see previous section. The LONG-NO-SNe at small redshift are clustered at low values of $T_{90}/(1+z)< 4$ (s) and with $L_{iso}> 6.29 \times 10^{48}$ erg $s^{-1}$ with the exception of GRB 061110A, which has $T_{90}/(1+z)=26$ s, thus showing a different behavior in terms of $T_{90}/(1+z)$ compared to the LONG-SNe category. 
Moreover, the time-averaged luminosity -$T_{90}/(1 + z)$ plane shows that all GRBs-SNe (A+B) types, which lie within 1$\sigma$ of the best-fit correlation line, also belong to the category of the engine-driven mechanism. For example, GRB 060218 is a bit off the 1$\sigma$ correlation and is classified a shock break out burst; however, if correction for the beamed luminosity is applied, GRB 060218 is not anymore an outlier. Thus, whether or not the driven engine mechanism can explain the tighter correlation is still an open question. Nevertheless, it may be interesting in a future study to investigate under which conditions the launching of a SN is possible contemporaneously with the occurrence of the Dainotti relation with that particular slope for the LONG-SNe sample.
 
\end{appendix}

\end{document}